\documentclass[prd,preprintnumbers,floatfix,aps,nofootinbib,notitlepage,showpacs,twocolumn
]{revtex4}
\usepackage{amsmath}
\usepackage{graphicx}
\usepackage{epstopdf}
\usepackage{latexsym}
\usepackage{amssymb}

\begin{document}

\title{3-form inflation in Randall-Sundrum II}
\author{Bruno J. Barros and Nelson J. Nunes}
\affiliation{Instituto de Astrof\'isica e Ci\^encias do Espa\c{c}o, Universidade de Lisboa, 
Faculdade de Ci\^encias, Campo Grande, PT1749-016 Lisboa, Portugal}
\date{\today}

\begin{abstract}
It has been shown in the last few years  that 3-form fields present viable cosmological solutions for inflation and dark energy with particular observable signatures distinct from those of canonical single scalar field inflation. The aim of this work is to explore the dynamics of a single 3-form in five dimensional  Randall-Sundrum II braneworld scenario, in which a 3-form is confined to the brane and only gravity propagates in the bulk. We compare the solutions with the standard four dimensional case already studied in the literature. In particular, we evaluate how the spectral index and the ratio of tensor to scalar perturbations are influenced by the presence of the bulk and put constraints on the parameters of the models in the light of the recent Planck 2015 data. 
\end{abstract}	

\pacs{98.80.-k,98.80.Jk}
\maketitle

\section{Introduction}
Primordial inflation provides solutions for cosmological puzzles such as the flatness and horizon problems and also explains the emergence of the primordial density fluctuations essential for the formation of the large scale structure observed today \cite{Guth:1980zm,Linde:1983gd}. Inflation is typically studied considering a self interacting  scalar field and has been widely studied in the literature (see \cite{Bassett:2005xm,Martin:2013tda} for reviews). 
The possibility of the energy source of the inflationary expansion
to be of a non-scalar nature has, however, never been excluded.
It is, therefore, important to understand the nature of higher spin fields and how robust they are
in order to fully test their applications in cosmology.
Inflation considering higher spinor fields has been investigated in the past and these
models are also important due to their connection to string theory
scenarios \cite{Frey:2002qc,Gubser:2000vg,Groh:2012tf}. Vector inflation has been studied in Ref.~\cite{Ford:1989me}, however, for inflation to proceed, the vector needs a nonminimal coupling and the model appears  to feature some instabilities. Inflation  with a 2-form field 
resembles much the vector inflation with the same problems \cite{Germani:2009iq,Koivisto:2009sd}. 

A 3-form has been shown to present viable solutions, not only for inflation \cite{Koivisto:2009ew,Koivisto:2009fb,Mulryne:2012ax,DeFelice:2012jt}, but also for describing dark energy  \cite{Koivisto:2012xm}.
Inflation driven by two 3-form fields has also been studied and does presents interesting results \cite{Kumar:2014oka}. 

 The natural question that arises now is how these properties translate to an extra-dimensional cosmological scenario. For example, in the Randall-Sundrum II model, proposed in 1999 \cite{Randall:1999vf}, our universe is confined to a four dimensional 3-brane, where the standard model particles reside, embedded in a five dimensional slice of an anti-de Sitter (AdS) space-time, the bulk. The presence of the bulk modifies the evolution equations \cite{Brax:2003fv}, more specifically, the Friedmann equation leads to a non-standard expansion law of the universe at high energies, while reproducing the standard four dimensional cosmology at low energies. One particular feature of the RSII model is that the tensor modes are enhanced due to the presence of the five dimensional bulk \cite{Langlois:2000ns,Langlois:2002bb}.  Chaotic inflation on the brane has been investigated in Ref. \cite{Maartens:1999hf} and it was shown that the inflationary predictions are modified from those in the four dimensional standard cosmology. Quintessential inflation from brane worlds has also been explored in \cite{Nunes:2002wz} and also inflation in the context of a Gauss-Bonnet brane cosmology \cite{Lidsey:2003sj}.
 More recently, simple inflationary models  in the context of  braneworld cosmology  were analysed against the 2015 Planck data \cite{Okada:2014eva,Okada:2015bra}.
 
 It is important to compare the dynamics of inflation with scalar fields with  the dynamics where higher order fields are considered.  The purpose of this work is, therefore, to study braneworld inflationary models driven by a single 3-form, confined to the brane, in the light of the Planck 2015 results \cite{Ade:2015lrj,Ade:2015xua}. In Sec. \ref{RSII} we introduce the 3-form model in the Randall Sundrum II braneworld. We follow to  rewrite the equations of motion in terms of a first order dynamical system for which we identify the critical points and analyse their stability for a specific form of the potential.  We explore the main differences of the dynamics compared with the four dimensional case. In Sec. \ref{perturbations} we write the power spectra for the scalar and tensor perturbations, calculate the cosmological parameters tensor to scalar ratio and spectral index and evaluate  how sensitive they are  to small changes in the brane tension.   We  find a lower bound on this parameter for a particular potential given the  recent Planck data \cite{Ade:2015lrj,Ade:2015xua}. Finally in Sec. \ref{conclusions} we summarize and discuss our results.

\section{3-form in Randall-Sundrum II}\label{RSII}

In the RSII scenario, our universe is confined to a single positive tension four dimensional 3-brane embedded in a five dimensional Anti de Sitter spacetime with a negative (bulk) cosmological constant.
A single 3-form field $A_{\mu\nu\rho}$ minimal coupled to Einstein gravity is confined to the brane,
\begin{eqnarray}
\label{action}
S &=& -\int d^5 x \sqrt{-g^{(5)}} \left( \frac{R}
{2\kappa_5^2} + \Lambda_5 \right) \nonumber \\
&-&  \int d^4 x \sqrt{-g^{(4)}} \left(\lambda -\frac{1}{48}F^2 -V(A^2)\right).
\end{eqnarray}
Here, $R$ is the Ricci scalar, $\Lambda_5$ is the bulk´s cosmological constant, $\lambda$ is the brane tension, $g^{(4)}$ and $g^{(5)}$ are the determinants of the four and five dimensional metrics, respectively. $\kappa^2=8\pi G$ and $F_{\alpha\beta\gamma\delta}$ is the Maxwell tensor given by,
\begin{equation}
F_{\alpha\beta\gamma\delta} = 4 \nabla_{[\alpha} A_{\beta\gamma\delta]},
\end{equation} 
 where square brackets denote antisymmetrization.

In order to avoid an excessive use of indices, we use the notation in which squaring means contracting all the indices, $A^2=A_{\mu\nu\rho} A^{\mu\nu\rho}$, and dotting means contracting the first index, $(\nabla \cdot A )_{\alpha\beta} = \nabla^{\mu} A_{\mu\alpha\beta}$.

We consider a Friedmann-Robertson-Walker Universe and take the scalar function $\chi (t)$ to parametrize the background contribution of the 3-form $A_{\mu\nu\rho}$. Thus,  the non-vanishing components are given by,
\begin{equation}
A_{ijk}=a^3 (t) \epsilon_{ijk} \chi(t),
\end{equation}
and therefore, $A^2=6\chi^2 (t)$, where $\epsilon_{ijk}$ is the standard Levi-Civita symbol and $i$,$j$ and $k$ denote spatial indices.

The action (\ref{action}) leads to the equations of motion for the 3-form,
\begin{equation}
\label{mot}
\nabla\cdot F= 12V'(A^2)A,
\end{equation}
and, due to antisymmetry, implies the additional set of constraints,
\begin{equation}
\nabla\cdot V'(A^2)A = 0.
\end{equation} 
The equations of motion in terms of the comoving field, $\chi$, are unmodified with respect to the previously studied four dimensional case because the matter fields are confined to the brane, 
\begin{equation}
\label{motion}
\ddot{\chi} + 3H\dot{\chi} + 3\dot{H}\chi + V_{,\chi} =0,
\end{equation}
where the third term is a new feature from the 3-form model, not present in the standard scalar field theory.
The generalization of the equations of motion to multiple 3-forms was done in Ref. \cite{Kumar:2014oka}. 

The presence of the bulk, however,  modifies Einstein's equations \cite{Brax:2003fv}. The five-dimensional Einstein's equations lead to the Friedmann equation,
\begin{equation}
H^2=\frac{\kappa ^2}{3} \rho\left[ 1 + \frac{\rho}{2 \lambda}\right] + \frac{\Lambda_4}{3} + \frac{\mu}{a^4},
\end{equation}
where  $\Lambda_4$ is the brane  four-dimensional cosmological constant and the last term represents the influence of the bulk gravitons on the brane. In what follows we will use units where $\kappa^2=1$ and we will assume that $\Lambda_4=\mu=0$, leaving us with,
\begin{equation}
\label{fridmannRSII}
H^2=\frac{1}{3} \rho\left[ 1 + \frac{\rho}{2 \lambda}\right].
\end{equation}

When we inspect Eq.~(\ref{fridmannRSII}), we note that the expansion rate is larger at high energies ($\rho \gg 2\lambda$), which means that the friction term in Eq.~(\ref{motion}) is larger in that regime. This means that the field $\chi(t)$ rolls slower and, for the same initial conditions,  inflation can last longer in this five-dimensions set up than in  the four-dimensional case. The Friedmann equation in the standard cosmology is reproduced in the limit of low energies, $\rho \ll 2\lambda$.

We can define the energy density and pressure for the field in the form,
\begin{eqnarray}
\rho &=& \frac{1}{2} (\dot{\chi} + 3H\chi)^2 + V,\label{energy} \\
p &=& -\frac{1}{2} (\dot{\chi} + 3H\chi)^2 -V + V_{,\chi}\chi.
\end{eqnarray}

\subsection{Dynamics of the 3-form on the brane}\label{dynamics}

In order to study the dynamics of the 3-form on the brane we introduce the dimensionless variables,
\begin{eqnarray}
x &\equiv&  \kappa \chi, \label{x}\\
y^2 &\equiv& \frac{V}{\rho} \label{y}, \\ 
w &\equiv& \frac{\dot{\chi} + 3 H \chi}{\sqrt{2\rho}}, \label{w}\\
\Theta &\equiv& \left( 1+ \frac{\rho}{2 \lambda} \right)^{-1/2}, \label{theta}
\end{eqnarray}
where $x$ represents the comoving field $\chi$, $y$ and $w$ are, respectively, the normalized potential and  kinetic energies and $\Theta$ represents the correction term in Eq.~(\ref{fridmannRSII}).  These variables are subject to the constraint, that follows from Eq.~(\ref{energy}),
\begin{equation}
\label{constrangimento}
w^2 + y^2 =1.
\end{equation}

Using Eqs.~(\ref{energy}), (\ref{x}), (\ref{w}) and (\ref{theta}), the modified Friedmann and Raychaudhuri equations can be written as,
\begin{eqnarray}
H^2 &=& \frac{1}{3} \frac{V}{(1-w^2)} \Theta^{-2}, \label{friedMod}\\
\dot{H} &=& -V_{,x} x \left(\Theta^{-2} - \frac{1}{2}\right). \label{rayMod}
\end{eqnarray}

Substituting for $\rho$ in Eq.~(\ref{constrangimento}) using Eqs.~(\ref{y}) and (\ref{theta}), we obtain the useful relation for $\Theta$ in terms of the $x$ and $w$ variables,
\begin{equation}
\Theta^2= \frac{1-w^2}{1-w^2 + \frac{V}{2\lambda}}.
\end{equation}

Next we follow to rewrite the equation of motion Eq.~(\ref{motion}) in terms of  a system of first order differential equations for the new variables such that,
\begin{eqnarray}
x' &=& 3 \left( \sqrt{\frac{2}{3}} \Theta w - x \right), \label{xeq}
\\
w' &=& \frac{3}{2} \frac{V_{,x}}{V} (1-w^2) \left[ xw - \Theta  \sqrt{\frac{2}{3}} \right], \label{weq}
\end{eqnarray}
where a prime means differentiating in respect to the number of e-folds $N=\ln a(t)$, so that $x'=dx/dN$. 
This system of equations closes as $\Theta$ depends only on $x$ and $w$.

We immediately note that at low energies ($\rho\ll 2\lambda$ and therefore, $\Theta\approx1$) we end up recovering the 
four-dimensional equations studied in Ref.~\cite{Kumar:2014oka} even though the variables were normalized to $H^2$ instead of $\rho$ as we do here.  We would like to see now, how the presence of this correction term, $\Theta$, affects the dynamics of the system in comparison with the evolution in the four-dimensional case.

\subsection{Critical points}\label{critpoints}

Let us assume for now that $\Theta$ evolves sufficiently slow such that we can take it to be a constant within a few $e$-folds. We will see later that this assumption is actually supported by the numerical solutions. We can then identify the {\it instantaneous} critical points  of the dynamical system established by Eqs.~(\ref{xeq}) and (\ref{weq}). These are shown in Table \ref{tabela}.
\begin{table}[ht]
\centering
\begin{tabular}{ c|c c c c }
   & $x$ & $w$ & $V_{,x}/V$ & Description \\ \hline 
  A & $\pm \sqrt{\frac{2}{3}}  \Theta$ & $\pm 1$ & any & kinetic domination\\ 
      B & $x_{\rm ext}$ & $\sqrt{\frac{3}{2}}\frac{1}{\Theta} x_{\rm ext}$ & 0 & potential extrema 
\end{tabular}
\caption{\label{tabela} Instantaneous critical points of the dynamical system.}
\end{table}

The critical points A do not exist for the standard scalar field models \cite{Copeland:1997et} and result from the extra $3 H \chi$ term in the equation of motion (\ref{motion}). One of the eigenvalues vanishes, hence, we cannot infer anything regarding its stability from the linear analysis without specifying the form of the potential. The critical point B corresponds to the value of the field at the extrema of the potential, therefore, its stability is strongly dependent on whether it corresponds to a minimum or a maximum of the potential. 

From the analysis of the critical points we can see that, in the five dimensional set up, the critical points have a dependence on the correction term $\Theta$. This means that as the energy decreases, the instantaneous critical points move along the phase space and approach the four dimensional case at low energies, $\Theta =1$. 

\begin{figure}[ht]
\includegraphics[width=8.5cm]{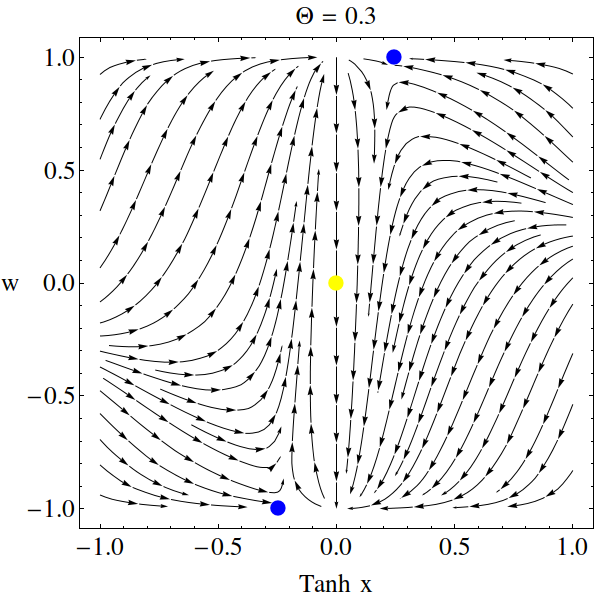}
\caption{\label{phase1}Phase space  $(\tanh (x),w)$ for $V=e^{\chi^2} -1$ at $\Theta=0.3$. } 
\end{figure}

\begin{figure}[ht]
\includegraphics[width=8.5cm]{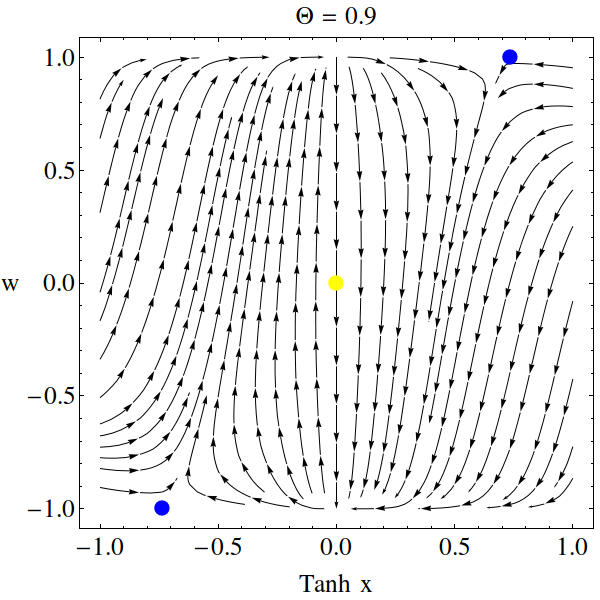}
\caption{\label{phase2}Phase space $(\tanh (x),w)$ for $V=e^{\chi^2} -1$ at $\Theta=0.9$. } 
\end{figure}

In Figs.~\ref{phase1} and \ref{phase2} is shown the phase space portrait for a potential of the form $V=e^{\chi^2} -1$. 
Comparing these figures we, again, note that the critical points A (upper and lower dots) are shifted along the $x$ axis as the system evolves and will eventually end at $x=\pm \sqrt{2/3}$ (4 dim case). As we will see in Sec.~\ref{inflation},  at the critical points A (top and bottom dots), the universe inflates and critical point B (central dots) corresponds to the attractor and potential minimum for this potential where reheating happens as usual \cite{DeFelice:2012wy}.

\begin{figure}[ht]
\includegraphics[width=8.5cm]{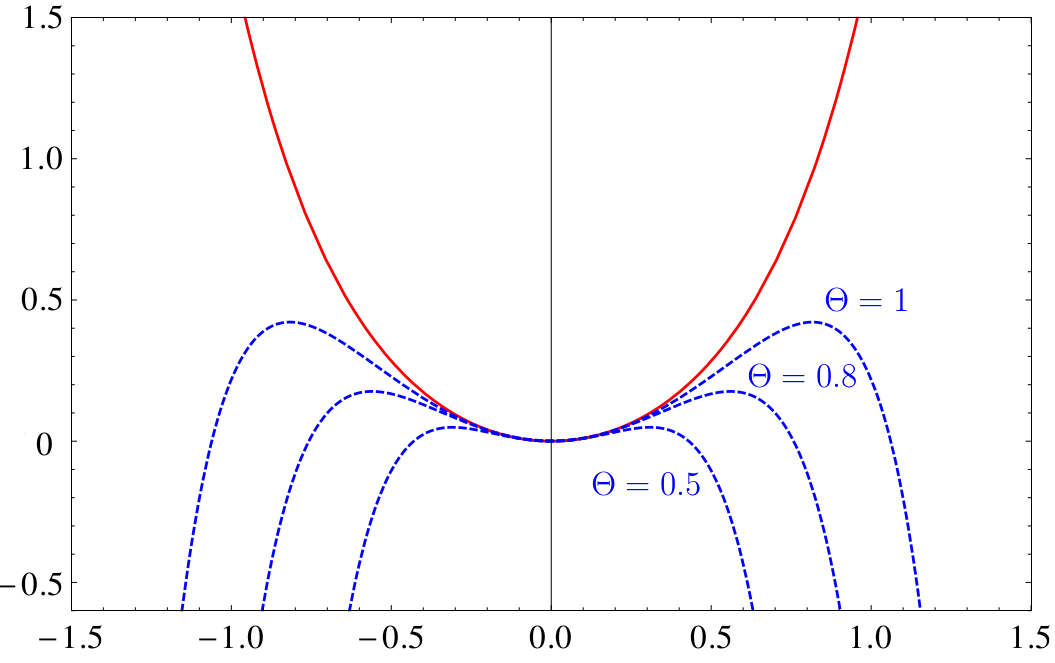}
\caption{\label{effective1} Potential $V(\chi)$ (solid line) and effective potential $V_{eff}$ (dashed lines) for the potential $V=e^{\chi^2} -1$ for different values of $\Theta$.} 
\end{figure}

\begin{figure}[ht]
\includegraphics[width=8.5cm]{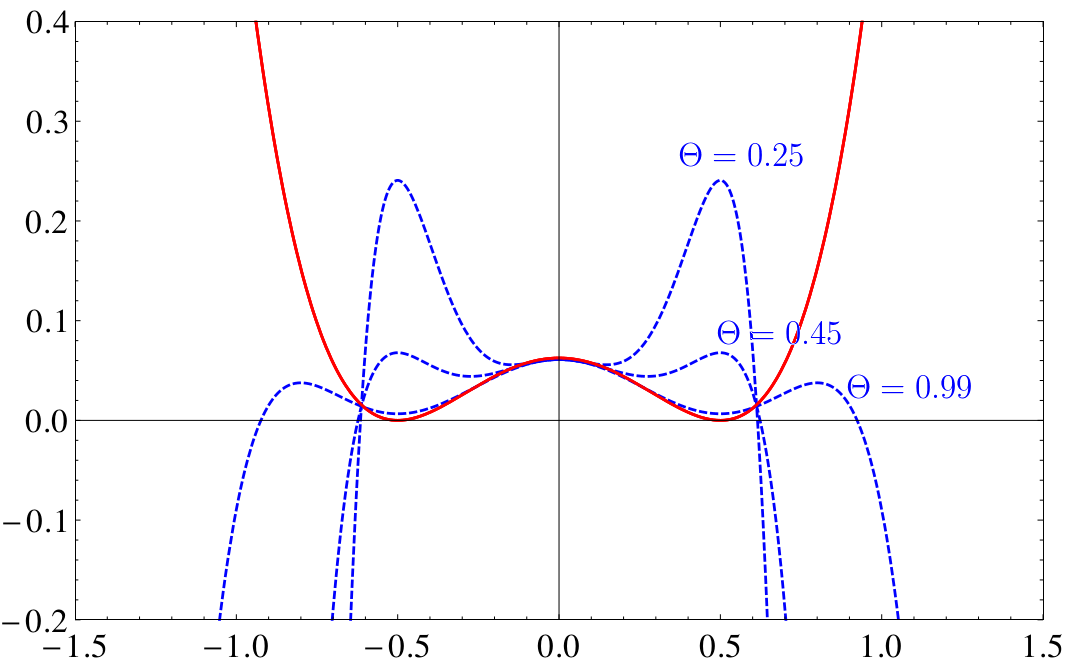}
\caption{\label{effmexican} Potential $V(\chi)$ (solid line) and effective potential $V_{eff}$ (dashed lines) for the potential $V=(\chi^2-0.5^2)^2$ for different values of $\Theta$.} 
\end{figure}

An alternative way to study the stability of the critical points is by defining the effective potential,
\begin{equation}
V_{{\rm eff},\chi}= 3 \dot{H}\chi + V_{,\chi}.
\end{equation}
We illustrate the potential and the corresponding effective potential for $V=e^{\chi^2} -1$ in Fig.~\ref{effective1}.
We can observe the shift in the value of the instant critical points as the energies decrease, i.e., as $\Theta$ approaches unity,  where  the critical points are $x = \pm \sqrt{\frac{2}{3}}$ as we can also verify in Table \ref{tabela}.
One interesting feature regarding the dynamics of a 3-form in RSII is that the $\Theta$ dependence of  the dynamics can change the stability of the critical points as the energy decreases.
For example, in Fig.~\ref{effmexican}, we traced the Landau-Ginzburg potential 
\begin{equation}
V(\chi)=(\chi^2-c^2)^2,
\end{equation}
with $c=0.5$ (solid), and its effective potential (dashed)  at different values of $\Theta$ and we observe that at early times the potential minima at $x = \pm 0.5$  are initially unstable and, as the  energy decreases, they become  stable.

\subsection{Initial conditions and slow roll regime}

In order to study inflation we need to understand  how the slow-roll parameters are modified in this set up. Analogously to the scalar field  as well as 3-forms \cite{Koivisto:2009ew,DeFelice:2012jt} the parameters are defined by $\epsilon \equiv -\dot{H} / H^2 = -d\ln H/ dN$ and $\eta=\epsilon ' / \epsilon - 2\epsilon$. One manner to establish a sufficient condition for inflation is, $\epsilon \ll 1$ and 
$|\eta|\ll 1$,
which must last for at least $\approx 50$ $e$-folds.  For our RSII model we have,
\begin{eqnarray}
\epsilon &=& \frac{3}{2} x \frac{V_{,x}}{V} (1-w^2) (2 - \Theta^2), \\
\eta &=& \frac{x'(V_{,x} + V_{,xx}x)}{V_{,x}x} + 6x \frac{V_{,x}}{V} (1-w^2) \frac{\Theta^2 -1}{2-\Theta^2},
\end{eqnarray}
where the terms in $\Theta$ signal the new contributions to the slow-roll parameters.

\subsection{3-form inflation on the brane}\label{inflation}

\begin{figure}[ht]
\includegraphics[width=8.5cm]{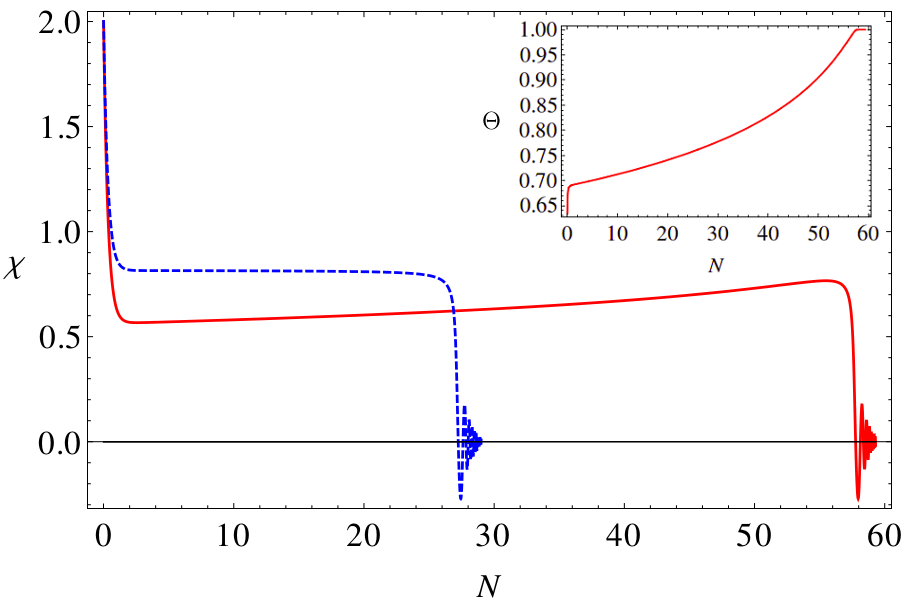}
\caption{\label{exp} Solutions for the system (\ref{xeq})--(\ref{weq}) for the four dimensional case (dashed line) i.e. for $\Theta=1$ already studied in \cite{Kumar:2014oka} and for the RSII model (solid line) when $\Theta$ is given by Eq.~(\ref{theta}) for $V=V_0 (e^{\chi^2} -1)$, $V_0=10^{-14}$, $\lambda=10^{-12}$ and for the initial conditions $(x_0,w_0)=(2,0.9055)$. The smaller panel shows the change in $\Theta$, for the RSII model, as the system evolves. } 
\end{figure}

\begin{figure}[ht]
\includegraphics[width=7.5cm]{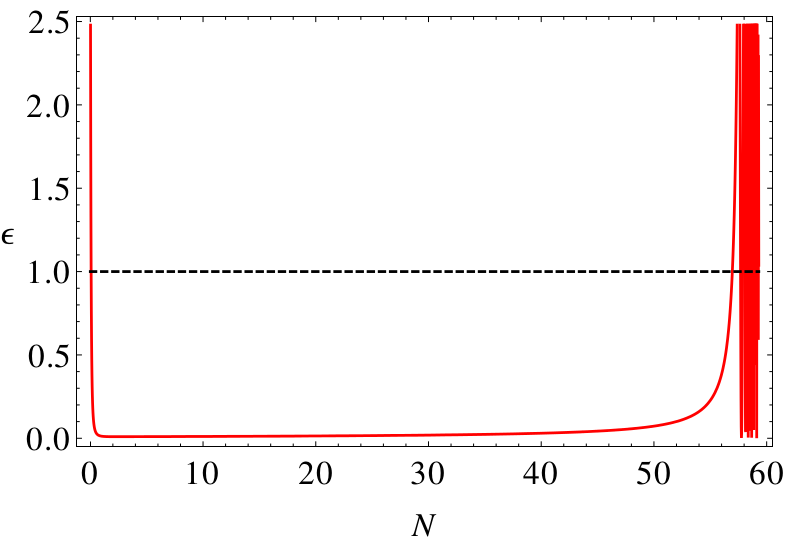}
\caption{\label{epsilon} Change in the slow roll parameter $\epsilon$ for the solutions for the system (\ref{xeq})--(\ref{weq}) for the RSII model for $V=V_0 (e^{\chi^2} -1)$, $V_0=10^{-14}$, $\lambda=10^{-12}$ and for the initial conditions $(x_0,w_0)=(2,0.9055)$. The dashed line marks $\epsilon=1$ just for reference.} 
\end{figure}

In this subsection we present inflationary solutions for the system (\ref{xeq})--(\ref{weq}).  We also compare the evolutions between the four and five dimensional cases.
Inspecting  Fig.~\ref{exp} and Fig.~\ref{epsilon} we note that inflation happens when the field is on the plateau of the evolution that for the four dimensional case is flat and corresponds to the critical point $\chi=\pm \sqrt{2/3}$ \cite{Kumar:2014oka}.  For the RSII case, however,  the plateau has a gentle slope due to the dependence of the instantaneous critical points on $\Theta$ (we saw that $\chi = \pm \sqrt{2/3}  \Theta$) up to the point in which  $\chi=\pm \sqrt{2/3}$. We can also note that, for the same initial conditions,  inflation lasts about 30 $e$-folds longer in the five dimensional set up due to the fact  that there is additional friction to the field's evolution. When inflation ends, the field goes to the attractor $\chi=0$ which is the potential minimum (critical point B in  Table \ref{tabela}).
%

\section{Cosmological perturbations}
\label{perturbations}

Since the 3-form is confined to the brane and neglecting any backreaction effects of the metric fluctuationsb in the fifth dimension  \cite{Maartens:1999hf}, the power spectrum of the curvature perturbations reads,
\begin{equation}
\label{power}
\mathcal{P}_{\zeta} = \left.\frac{2 H^4}{m_{\rm pl}^2 \pi V_{,\chi} \chi c_s} \right|_*,
\end{equation}
where, 
 $*$ indicates horizon crossing $ c_s k=aH$, and the sound speed  is given by \cite{Koivisto:2009fb,Mulryne:2012ax}, 
\begin{equation}
c_s^2 = \frac{V_{,\chi\chi} \chi}{V_{,\chi}}.
\end{equation}
From the Planck 2015 results \cite{Ade:2015xua}, we fix the power spectrum of scalar perturbations as $\mathcal{P}_{\zeta}(k_0) = 2.196 \times 10^{-9}$ for the pivot scale chosen at $k_0 = 0.002$ Mpc$^{-1}$.

The spectral index is given by
\begin{equation}
\label{spectral}
n_s -1 = -5\epsilon - \frac{\dot{c}_s}{c_s H} - \epsilon c_s^2 + \frac{V_{,\chi}}{3\chi H^2} (1+c_s^2),
\end{equation}
which, as the power spectrum, also has a dependence on the speed of sound.

In the Randall-Sundrum model, however, the amplitude of the tensor modes are modified and the respective power spectrum reads \cite{Langlois:2000ns},
\begin{equation}
\label{at}
\mathcal{P}_T = \frac{64\pi}{m_{\rm pl}^2} \left( \frac{H}{2\pi} \right)^2 F^2(x_0) |_*,
\end{equation}
where $F$ is a correction function, 
\begin{equation}
\label{f}
F(x)= \left[ \sqrt{1+x^2} - x^2 \ln \left( \frac{1}{x} + \sqrt{1+ \frac{1}{x^2}} \right) \right]^{-1/2}, 
\end{equation}
and
\begin{equation}
x_0 = \left(\frac{3}{4\pi\lambda} \right)^{1/2} H M_{\rm Pl}.
\end{equation}
For $x_0 \ll 1$, $F(x_0) \simeq 1$ and Eq.~(\ref{at}) reduces to the standard cosmology formula, and for $x_0 \gg 1$, $F(x_0) \simeq \sqrt{3x_0  /2}$. Finally, the tensor to scalar ratio is then,
\begin{equation}
\label{tsratio}
r\equiv \frac{\mathcal{P}_T}{\mathcal{P}_{\zeta}} = \frac{8}{H^2} V_{,\chi} \chi c_s F^2 (x_0).
\end{equation}

\begin{figure}[ht]
\includegraphics[width=8.5cm]{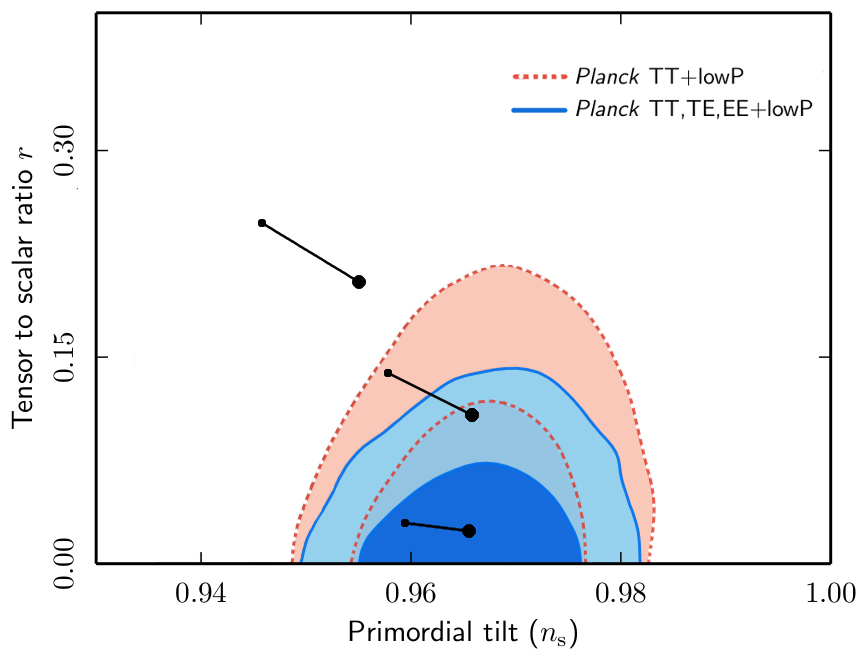}
\caption{\label{results}  Comparison of the spectral index and the tensor to scalar ratio against the recent Planck 2015 data \cite{Ade:2015lrj} for 50 (small dot) and 60 (large dot) $e$-folds for different values of the brane tension $\lambda$. We considered the potential  in Eq.~(\ref{pot}) with $b=-0.245$. The bars represent, from bottom to top, the solutions with $\lambda=10^{-5}$,  $\lambda = 3 \times 10^{-9}$ and $\lambda =10^{-10}$ in units 
$\kappa^2=1$).} 
\end{figure}

We are now ready to compare the cosmological parameters, scalar to tensor ratio and spectral index, of our inflationary setting with the 2015 Planck data \cite{Ade:2015lrj}.
First we consider a form of the scalar potential which has been proven in Ref. \cite{Kumar:2014oka}  to lead to a viable cosmology in the four dimensional set up (although for a two 3-form system) and to produce a good fit to the Planck 2013 results,
\begin{equation}
\label{pot}
V=V_0 (\chi^2 + b\chi^4),
\end{equation}
where $V_0$ and $b$ are free parameters.
In Fig.~\ref{results} the bottom bar represents the prediction for the five dimensional case with 
$\lambda=10^{-5}$. With this value of the brane tension, the evolution quickly reaches  $\Theta \approx 1$  which means that this case is practically indistinguishable from the four dimensional solution. 
When we lower the brane tension and consequently increase the five dimensional effects, we observe that the predictions worsen due to the presence of the correction $F^2(x_0)$ in Eq.~(\ref{at}), which enhances the tensor to scalar ratio. For  $\lambda =10^{-10}$, corresponding to $\lambda \simeq (3.9 \times 10^{16}\,\,{\rm GeV})^4$ (corresponding to the upper bar) the predictions are beyond the Planck TT+lowP contour limits. We find a lower bound, for 60 $e$-folds, of $\lambda \simeq 1.5 \times 10^{-9}$, corresponding to $\lambda \geq (7.6\times 10^{16}\,\,{\rm GeV})^4$, for the inflationary predictions to be within the Planck TT,TE,EE+lowP contour limits.

\begin{figure}[ht]
\includegraphics[width=7.5cm]{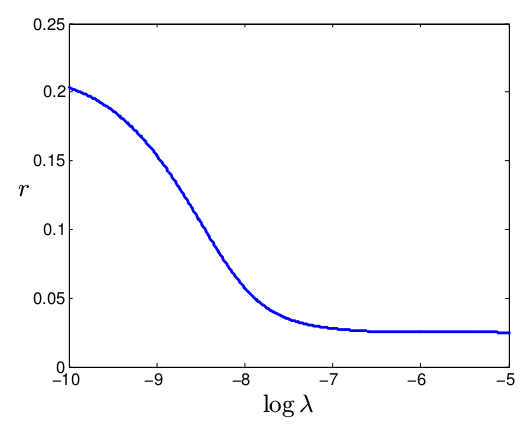}
\caption{\label{r1} $\log \lambda$ vs $r$, for the potential (\ref{pot}), with $b=-0.245$, for 60 $e$-folds, for different values of the brane tension $\lambda$. } 
\end{figure}

\begin{figure}[ht]
\includegraphics[width=7.5cm]{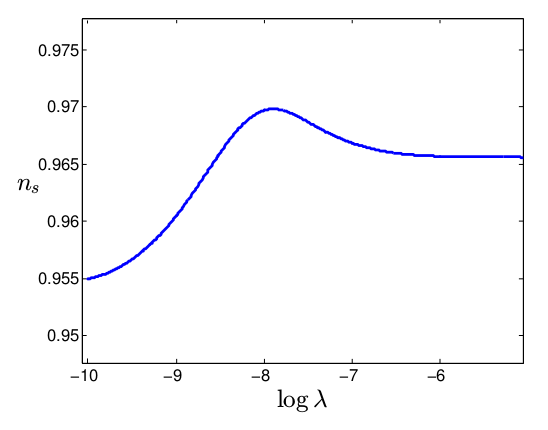}
\caption{\label{ns} $\log \lambda$ vs $n_s$, for the potential (\ref{pot}), with $b=-0.245$, for 60 $e$-folds, for different values of the brane tension $\lambda$. } 
\end{figure}

In Fig.~\ref{r1} we analyse how the brane tension and the tensor to scalar ratio are related  as $\lambda$ is lowered for 60 $e$-folds. For $\lambda < 10^{-7}$, $r$ quickly increases due to the presence of $F^2$ in Eq.~(\ref{tsratio}), making the predictions worse as we also saw in Fig.~\ref{results}. In Fig.~\ref{ns} we present the relation between the spectral index and the logarithm of the brane tension $\lambda$. As expected, $n_s$ is almost insensitive to $\lambda$ for large values of this quantity. This is the case because 
at large $\lambda$ the standard scenario is recovered and as in the scalar picture of the three-form the scalar potential is quadratic, the spectral index must be close to $n_s \sim 0.967$ \cite{Mulryne:2012ax}.

\begin{figure}[ht]
\includegraphics[width=7.5cm]{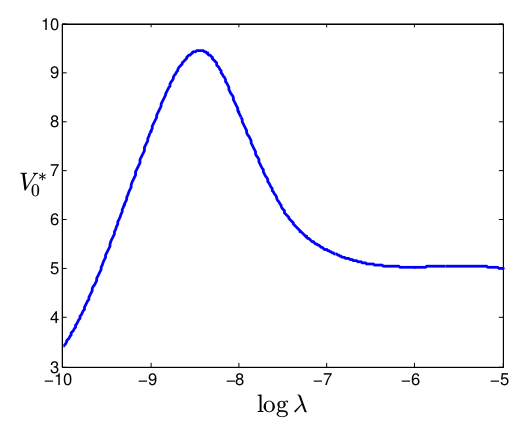}
\caption{\label{vzero} $\log \lambda$ vs $V_0^* = V_0 \times 10^{12}$, for the potential (\ref{pot}), with $b=-0.245$, for 60 $e$-folds, for different values of the brane tension $\lambda$. } 
\end{figure}

When we lower the brane tension, in order to keep the power spectrum of scalar perturbations fixed as $\mathcal{P}_{\zeta}(k_0) = 2.196 \times 10^{-9}$, for the pivot scale chosen at $k_0 = 0.002$ Mpc$^{-1}$, we also have to change $V_0$ in order to 
ensure this normalization.  This relation is shown in Fig.~\ref{vzero}.

\section{Summary and discussion}
\label{conclusions}

In this work we explored the main differences between the dynamics of a single 3-form in the Randall-Sundrum II braneworld and the standard four dimensional case \cite{Koivisto:2009ew}. We followed to write the equations of motion for the 3-form model in terms of a system of first order differential equations (\ref{xeq})--(\ref{weq}).  By defining a set of useful variables $(x,y,w,\Theta)$  we identified what we called the instantaneous critical points which now have a dependence on the correction term, $\Theta$, arising from the modified Friedmann equation.   
We illustrated the effects that take place at high energies by showing the phase space of the system  at different stages of the universe, or in other words, for different values of $\Theta$, and by interpreting them as a modification to the effective potential. 
It was observed that in five dimensions the stability of some instantaneous critical points can change with the energy. 
We presented an inflationary solution for the potential in Eq.~(\ref{pot}) and computed the respective
tensor to scalar ratio (\ref{tsratio}) and spectral index (\ref{spectral}). 
We were able to fit the cosmological predictions with the recent Planck 2015 data \cite{Ade:2015lrj} for a choice of parameters and saw that the effects of the braneworld bring the observables away from the central region of the data contours. By performing this study, we found a lower bound for the brane tension for the potential (\ref{pot}) such that the observables' values remain inside the contours of the Planck TT,TE,EE+lowP.

\begin{acknowledgments}
The authors thank Carsten van de Bruck and Tomi Koivisto for comments on the manuscript. 
N.J.N was supported by the Funda\c{c}\~{a}o para a Ci\^{e}ncia e Tecnologia 
(FCT) through the grants EXPL/FIS-AST/1608/2013 and UID/FIS/04434/2013. 
\end{acknowledgments}

\end{document}